# Electrosteric enhanced stability of functional sub-10 nm cerium and iron oxide particles in cell culture medium

B. Chanteau[a], J. Fresnais[b] and J.-F. Berret[a,@]

*(a)* : *Matière et Systèmes Complexes, UMR 7057 CNRS Université Denis Diderot Paris-VII, Bâtiment Condorcet, 10 rue Alice Domon et Léonie Duquet, 75205 Paris, Fra*nc*e* – *(b)* : *UPMC Univ Paris 06 - Laboratoire Liquides Ioniques et Interfaces Chargées, UMR 7612 CNRS, 4 place Jussieu-case 63, F-75252 Paris Cedex 05 France*

**Abstract** : Applications of nanoparticles in biology require that the nanoparticles remain stable in solutions containing high concentrations of proteins and salts, as well as in cell culture media. In this work, we developed simple protocols for the coating of sub-10 nm nanoparticles and evaluated the colloidal stability of dispersions in various environments. Ligands (citric acid), oligomers (phosphonate-terminated poly(ethylene oxide)) and polymers (poly(acrylic acid)) were used as nanometer-thick adlayers for cerium ($CeO_2$) and iron ($\gamma$-$Fe_2O_3$) oxide nanoparticles. The organic functionalities were adsorbed on the particle surfaces via physical (electrostatic) forces. Stability assays at high ionic strength and in cell culture media were performed by static and dynamic light scattering. Among the three coating examined, we found that only poly(acrylic acid) fully preserved the dispersion stability on the long term (> weeks). The improved stability was explained by the multi-point attachments of the chains onto the particle surface, and by the adlayer-mediated electrosteric interactions. These results suggest that anionically charged polymers represent an effective alternative to conventional coating agents.

## I – Introduction

Inorganic nanoparticles (NP) made from gold, metal oxides or semiconductors are emerging as the central constituents of future nanotechnological developments. Interest stems from the combination of complementary attributes, such as a size in the nanometer range and unique physical features, including high reactivity, magnetic or optical properties. In biomedicine, inorganic nanoparticles have attracted interest with respect to applications such as magnetic separation, biosensor devices, diagnostics, imaging and therapeutics [1]. Nanomaterials possess also very high specific surfaces (a few hundred of square meters per gram of 10 nm particles) which may enhance their chemical reactivity with respect to living organisms. As compared to conventional chemicals, the risk assessment of nanomaterials towards human health has not been fully appreciated. Issues dealing with the impacts, fate and toxicity of nanoparticles on the environment have become the focus of many recent researches, and in particular of in vitro toxicology experiments [2-12]. The objectives of toxicology assays, such as the MTT, neutral red or WST1 are the quantitative determination of the viability of living cells that were incubated with nanomaterials [13]. Brunner et al. submitted rodent fibroblasts to various industrially important particles and measured the cell responses in terms of metabolic activity and cell proliferation [8]. Their results were that most uncoated particles were cytotoxic. Limbach *et al.* focused on the transport and uptake of cerium oxide NPs into human lung fibroblasts, and found that the size of the particles was the decisive criterion [4]. In these studies and in others [2,5,7,9,14-16], it was also recognized that the physico-chemical characteristics of the particles (and not only the nature of their atomistic constituents) played a crucial role on the cellular uptake. The chemistry of the interfaces between nanocrystal and solvent was also anticipated to be a key feature of the cell/NP interactions. Should the particle surfaces be charged or neutral, should they be tethered with polymers, oligomers or low-molecular weight molecules, or should the particles be aggregated into large clusters, the interactions towards cells would be drastically different [2-4,6-10,12,17].

In the present paper, we emphasize a feature that was not systematically addressed in toxicology studies of nanomaterials, namely their colloidal stability in a cellular culture medium. In a certain number of reports, it was found that in culture media particles aggregated almost systematically, yielding the formation of large clusters and sedimentation [7,10,12,14,16,18,19]. In addition, the aggregation was irreversible, and in some cases corresponded to the loss of the physical size-related attributes [10]. Other scenarios have shown that in serum rich media, particles were covered by proteins such as immunoglobulins, yielding large core-corona hybrid structures [20]. In this work, we have investigated the role of the coating on the colloidal stability of cerium ($CeO_2$) and iron oxide ($\gamma$-$Fe_2O_3$) nanoparticles in complex environments. The particles have been dispersed in the culture medium that is employed for the growth of eukaryotic cells (Dubbelco's Modified Eagle's Medium supplemented with calf serum and antibiotics). Three different coating using small organic molecules and macromolecules were produced through simple protocols and physical (electrostatic) interactions with the particles [21]. Here, we demonstrate that electrosteric (a combination of electrostatic and steric repulsions [22]) coating can be robust and efficient for stabilizing nanoparticles.

## II – Experimental

### II.1 – Nanoparticles

*Cerium oxide nanoparticles (nanoceria)* : The cerium oxide nanoparticles (bulk mass density [23] $\rho = 7.1$ g cm$^{-3}$) were synthesized in nitric acid solution at pH 1.4 by thermohydrolysis of a cerium(IV) nitrate salt solution at high temperature. The thermohydrolysis resulted in the homogeneous precipitation of cerium oxide nanoparticles [24]. The size of the particles was controlled by addition of hydroxide ions during the procedure. An image of transmission electron microscopy (TEM) obtained from a dispersion at 0.2 wt. % is illustrated in Fig. 1a. It shows that the nanoceria consisted of isotropic agglomerates of ~ 2 nm crystallites with faceted morphologies. An image analysis of the TEM data allowed us to derive the size distribution of the NPs. It was found to be well-accounted for by a log-normal function with median diameter $D_0 = 7.0$ nm and polydispersity $s = 0.15$ [25]. The polydispersity was defined as the ratio between the standard deviation and the average diameter. Wide-angle x-ray scattering confirmed the crystalline fluorite face-centered cubic structure of the nanocrystallites (see Supporting Informations, Fig. SI.1). Using static and dynamic light scattering, the molecular weight and the hydrodynamic diameter of the particles in acidic conditions were determined at $M_W = 330$ kg mol$^{-1}$ and at $D_H = 9$ nm, respectively (Table I) [26]. As synthesized, the cerium oxide nanosols were stabilized by electrostatic repulsive interactions. An increase in pH or ionic strength produced however an instantaneous and irreversible aggregation of the particles, and ultimately a destabilization of the sols [23,24].

*Iron oxide nanoparticles (maghemite)* : The iron oxide nanoparticles (bulk mass density $\rho = 5100$ kg m$^{-3}$) were synthesized according to the Massart method [27] by *i)* alkaline co-precipitation of iron(II) and iron(III) salts, *ii)* oxidation of the magnetite (Fe$_3$O$_4$) into maghemite ($\gamma$-Fe$_2$O$_3$) NPs, and *iii)* size-sorting by subsequent phase separations [27,28]. At the end of the process, the particles were at the concentrations $c \sim 10$ wt. % and in acidic conditions (pH 1.8). As for the nanoceria, the particles were positively charged, with nitrate counterions adsorbed on their surfaces. The resulting electrostatic repulsion between nanocolloids also insured a remarkable colloidal stability of the dispersions, typically over several years. As for nanoceria, the size distributions as determined from TEM measurements (Fig. 1b and 1c) could be represented by a log-normal function, with median diameter $D_0$ and polydispersity $s$ [29]. For the present study, two maghemite batches of different particle sizes were synthesized, one in acidic condition at $D_0 = 7.1$ nm (batch referred to as "acidic" in Table I) and one coated with citrate counterions at $D_0 = 8.5$ nm (batch referred to "citrated"). For these systems, the polydispersity was estimated at $s = 0.26$ and $s = 0.29$, respectively. The magnetic nanosols were further characterized by electron microdiffraction, vibrating sample magnetometry, magnetic sedimentation and light scattering (see Supporting Informations for details) [29]. For iron oxide NPs, it should be noted that the hydrodynamic sizes appeared larger than the median diameters found by TEM. These differences were attributed to two effects [29] : the slight anisotropy of the $\gamma$-Fe$_2$O$_3$ particles (aspect ratio 1.2) and the fact that light scattering is more sensitive to the large particles of a distribution [30].

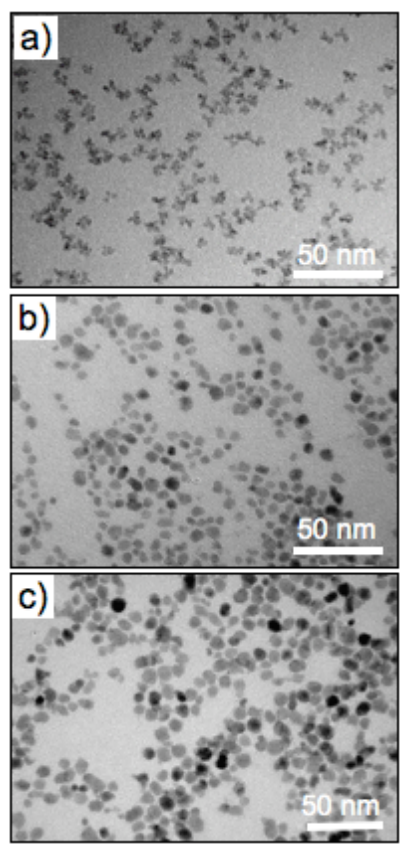

**Figure 1** : Transmission electron microscopy images (magnification ×120000) of the nanoparticles investigated in this work : a) cerium oxide (CeO$_2$), b) iron oxide ($\gamma$-Fe$_2$O$_3$) in acidic conditions and c) iron oxide ($\gamma$-Fe$_2$O$_3$) coated with citrate counterions at neutral pH. Image analysis of the TEM data allowed us to derive the size distribution of the different nanocrystals (Table I).

### II.2 – Chemicals and Coating

For this work, we have developed different types of coating based on the electrostatic adsorption of organic adlayer on the particles. Ligands (citric acid) [24], oligomers (phosphonate terminated poly(ethylene oxide) [26] and polymers (poly(acrylic acid)) [24] were used alternatively.

| nanoparticles | $D_0$ nm | s | $M_W$ kg mol$^{-1}$ | $D_H$ nm |
|---|---|---|---|---|
| CeO$_2$ | 7.0 | 0.15 | 330 | 9 |
| γ-Fe$_2$O$_3$ (acidic) | 7.1 | 0.26 | 1400 | 14 |
| γ-Fe$_2$O$_3$ (citrated) | 8.5 | 0.29 | 3000 | 23 |

**Table I** : Median diameter ($D_0$), polydispersity (s), molecular weight ($M_W$) and hydrodynamic diameter $D_H$ of the bare nanoparticles studied in this work. The size distributions of the particles were assumed to be log-normal for the three batches.

*Citric acid* : Citric acid is a weak triacid of molecular weight $M_W$ = 192.1 g mol$^{-1}$, which has three acidity constants at $pK_{A1}$ = 3.1, $pK_{A2}$ = 4.8 and $pK_{A3}$ = 6.4. Complexation of the surface charges with citric acid (Sigma Aldrich) was performed during the synthesis. It allowed to reverse the surface charge of the particles from cationic at low pH to anionic at high pH, through a ionization of the carboxyl groups. At pH 8, the particles were stabilized by electrostatic interactions mediated by the anionically charged ligands [24]. As a ligand, citrate ions were characterized by adsorption isotherms, *i.e.* the adsorbed species were in equilibrium with free citrates molecules dispersed in the bulk. The concentration of free citrates in the bulk was kept at the value of 8 mM [23,31,32], both in water and in culture medium. It should be noticed that the hydrodynamic diameter of the bare and citrate-coated particles were identical within the experimental accuracy, indicating a layer thickness inferior to 1 nm (Table II). The citrate-coated particles are denoted Cit–CeO$_2$ and Cit–γ-Fe$_2$O$_3$ in the manuscript.

*Phosphonate-PEG (PPEG)* : Poly(oxy-1 2 ethanediyl) α-(3-phosphonopropyl) ω-hydroxyl, abbreviated PPEG in the following is an oligomer of the phosphonated poly(oxyalkenes) class, that was produced by Rhodia Chemicals for coating purpose. PPEG titration curves have shown the presence of two $pK_{AS}$ ($pK_{A1}$ = 2.7 and $pK_{A2}$ = 7.8) associated with the ionization of the phosphonate group. Mixed solutions of nanoparticles and PPEG were prepared by simple mixing of dilute solutions prepared at pH 2 [26]. This ensured that no aggregation of nanoparticles occurred due to pH or salinity gap. The relative amount of each component was monitored by the volume ratio between the NP and oligomer stock solutions. This mixing ratio was chosen so that there was a large excess of oligomers per particle. Ammonium hydroxide (NH$_4$OH) was used to adjust the pH of PPEG–CeO$_2$ and PPEG–γ-Fe$_2$O$_3$ dispersions in the range of 1.5 to 10. In this range, the hydrodynamic diameter was found to be constant and at a value that exceeded that of the bare NPs by 2$h$ = 3 nm, $h$ being the adlayer thickness (Table II). A recent study of PPEG-coated nanoceria by light and neutron scattering revealed an average of 270 PPEG oligomers per particle for the dialyzed dispersions [26]. This value corresponds to density of adsorbed species of 1 nm$^{-2}$. Thanks to the external PEG-terminus, the particles were found to be neutral, as revealed by the very small values of the electrophoretic mobility.

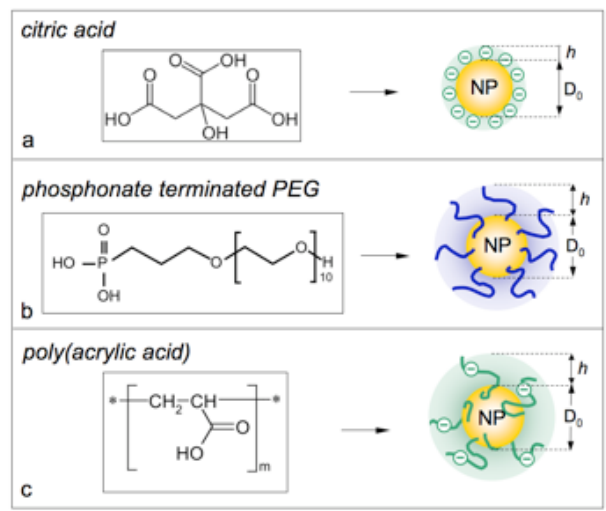

**Figure 2** : Schematic description of the three different coating developed in this study. The organic functionalities were adsorbed on the particle surfaces through electrostatic complexation. In cases a) and c), the coating was anionic whereas in case b) it was neutral. $D_0$ denotes the diameter of the particles as derived from TEM measurements. The thickness of the organic adlayers, noted $h$ was derived from dynamic light scattering. $h$ was of the order of a few angstroms for the citric acid, and a few nanometers for the others (see table II).

*Poly(acrylic acid)* : During the last years, poly(acrylic acid) (PAA) was frequently used as a coating agent of inorganic particles [15,17,24,25,33-35]. Poly(sodium acrylate), the salt form of PAA with a molecular weight $M_W$ = 2000 and 5000 g mol$^{-1}$ and a polydispersity of 1.7 was purchased from Sigma Aldrich and used without further purification. In order to adsorb polyelectrolytes onto the surface of the nanoparticles, we have followed the precipitation-redispersion protocol, as described elsewhere [24,29]. This simple technique allowed to produce large quantities of coated particles (> 1 g of oxides) within a relatively short time (<1 hour). The precipitation of the cationic cerium or iron oxide dispersion by PAA was performed in acidic conditions (pH 2). The precipitate was separated from the solution by centrifugation, and its pH was increased by addition of ammonium hydroxide. The precipitate redispersed spontaneously at pH 7 - 8, yielding a clear solution that now contained the polymer coated particles. The hydrodynamic sizes of PAA$_{2K}$–CeO$_2$, PAA$_{2K}$–γ-Fe$_2$O$_3$ and PAA$_{5K}$–γ-Fe$_2$O$_3$ were found to be D$_H$ = 13, 19 and 22 nm, respectively. These values were 4 - 5 and 8 nm larger than the hydrodynamic diameter of the uncoated particles, indicating a corona thickness $h$ = 2 - 2.5 nm for PAA$_{2K}$ and 4 nm for PAA$_{5K}$ (Table II). In terms of coverage, the number of adsorbed chains per particle was estimated in the range 50 – 70 [24,36]. As for the citrate-coated particles, it was checked by electrokinetic measurements that the PAA-coating resulted in strongly anionic charged interfaces [24,25,37]. Values of the electrophoretic mobilities were listed in the Supporting Informations section.

As a final step of the three above procedures, the dispersions were all dialyzed against DI-water which pH was first adjusted to 8 (Spectra Por 2 dialysis membrane with MWCO 12 kD). At this pH, 90 % of the carboxylate groups of the citrate and PAA coating were ionized. The schematic representations of the different coating are illustrated in Fig. 2. The suspension pH was adjusted with reagent-grade nitric acid (HNO$_3$) and with sodium or ammonium hydroxides. For the assessment of the stability with respect to the ionic strength, sodium and ammonium chloride (NaCl and NH$_4$Cl, Fluka) were used in the range $I_S$ = 0 – 1 M.

| Coated NP | R(c)/c cm$^{-1}$ | D$_H$ nm | 2$h$ nm |
|---|---|---|---|
| Cit–CeO$_2$ | 0.091 | 9 | < 1 |
| PPEG–CeO$_2$ | 0.076 | 12 | 3 |
| PAA$_{2K}$–CeO$_2$ | 0.150 | 13 | 4 |
| Cit–γ-Fe$_2$O$_3$[a] | 0.70 | 23 | < 1 |
| PPEG–γ-Fe$_2$O$_3$[b] | 0.51 | 17 | 3 |
| PAA$_{2K}$–γ-Fe$_2$O$_3$[b] | 0.60 | 19 | 5 |
| PAA$_{5K}$–γ-Fe$_2$O$_3$[b] | 0.71 | 22 | 8 |

**Table II** : Light scattering data found for the coated nanoparticles dispersed in de-ionized water. R(c) denotes the excess Rayleigh ratio, D$_H$ the hydrodynamic diameter and 2$h$ the increase in diameter due to the organic adlayer. In the concentration range explored (c = 10$^{-3}$ - 10$^{-1}$ wt. %), the Rayleigh ratios of the dispersions were found to vary linearly with the concentration, with slopes indicated in the first column. [a] : batch of maghemite particles coated during the synthesis, and referred to as "citrated" in Table I. [b] : batch of uncoated maghemite particles, and referred to as "acidic" in Table I.

II.3 – Culture medium

The particle stability was evaluated using a cell culture medium consisting of Dubbelco's Modified Eagle's Medium (DMEM, PAA Laboratories, GmbH), 10 % of fetal calf serum (FCS, PAA Laboratories, GmbH), 100 units l$^{-1}$ of penicillin and 100 mg l$^{-1}$ of streptomycin (PAA Laboratories, GmbH). In its commercial formulation, DMEM also contained 4.5 g l$^{-1}$ of D-Glucose, inorganic salts such as sodium chloride (6400 mg l$^{-1}$), sodium hydrogen carbonate (3700 mg l$^{-1}$), potassium and calcium chlorides (400 and 200 mg l$^{-1}$ respectively), as well as 17 amino acids in different concentrations ranging from 50 to 1000 mg l$^{-1}$ (e.g. L-Glutamine). Vitamins such as myo-inositol (7.2 mg l$^{-1}$) and folic acid (4 mg l$^{-1}$) were also present in the culture medium, however in a lesser amount. Most of these additives were charged species, and as a result increased the ionic strength of the solution up to $I_S$ = 0.16 M.

II.4 – Transmission Electron Microscopy

TEM experiments were carried out on a Jeol-100 CX microscope at the SIARE facility of University Pierre et Marie Curie (Paris 6). The TEM images of the cerium and iron oxide nanoparticles were obtained at a magnification ×120000 (Figs. 1) and their size distributions were derived using the ImageJ software (http://rsb.info.nih.gov/ij/). The diameter and polydispersity for CeO$_2$ were in good agreement with those determined by cryogenic transmission electron microscopy (cryo-TEM) in an earlier report [25].

II.5 – Static and Dynamic Light Scattering

Static and dynamic light scattering were performed on a Brookhaven spectrometer (BI-9000AT autocorrelator, λ = 632.8 nm) for measurements of the Rayleigh ratio $\mathcal{R}(q,c)$ and of the collective diffusion constant D(c). The Rayleigh ratio was obtained from the scattered intensity I(q,c) measured at the wave-vector q according to $\mathcal{R}(q,c) = \mathcal{R}_{std}(I(q,c) - I_S)/I_{Tol}$ where $\mathcal{R}_{std}$ is a standard Rayleigh ratio for toluene, $I_S$ and $I_{Tol}$ the intensities measured for the solvent and for the toluene and $q = \frac{4\pi n}{\lambda}\sin(\theta/2)$ (with n the refractive index of the solution and θ the scattering angle). In this study, the suspending solvent was de-ionized water with or without added salt, or the DMEM-based cell culture medium. In the regime of weak colloidal interactions, the Rayleigh ratio is predicted to follow a wave-vector and concentration dependence such as [30] :

$$\frac{Kc}{\mathcal{R}(q,c)} = \frac{1}{M_{w,app}}\left(1 + \frac{q^2 R_G^2}{3}\right) + 2A_2 c \qquad (1)$$

In Eq. 1, K = $4\pi^2 n^2 (dn/dc)^2/N_A \lambda^4$ is the scattering contrast coefficient ($N_A$ is the Avogadro number), the apparent molecular weight and $A_2$ is the second virial coefficient. The concentrations targeted in the present work were chosen so as to cover the concentration ranges classically investigated in toxicology assays.

Expressed in molar concentrations of metallic atoms, these concentrations ranged from $\{[Ce],[Fe]\} = 0.1 – 10$ mM. 10 mM of cerium (resp. iron) corresponded to a weight concentration of 0.17 wt. % (resp. 0.08 wt. %). In the dilute regime, $qR_G \ll 1$ and $A_2 \sim 0$ in Eq. 1, which then reads :

$$\mathcal{R}(q,c) = K M_{w,\mathrm{app}} c \qquad (2)$$

This latter equation emphasizes the fact that for sub-10 nm particles, the Rayleigh ratio does not depend on the wave-vector in the characteristic q-window of light scattering. In this regime, the coated NP dispersions were found to obey Eq. 2. Table II summarizes the experimentally determined Rayleigh ratio increments $\mathcal{R}(c)/c$ for the 7 systems considered here ($\theta = 90°$). In this work, the intensity scattered by nanoparticles dispersed in a cell growth medium was evaluated as a function of time and concentration. In case of a destabilization of the sol, the scattering intensity was expected to grow above the predictions of Eq. 2. With an accuracy better than 5 %, this technique was very sensitive to the state of a dispersions. With light scattering operating in dynamical mode, the collective diffusion coefficient D was determined from the second-order autocorrelation function of the scattered light. From the value of the coefficient, the hydrodynamic diameter of the colloids was calculated according to the Stokes-Einstein relation, $D_H = k_B T/3\pi\eta_S D$, where $k_B$ is the Boltzmann constant, T the temperature (T = 298 K) and $\eta_S$ the solvent viscosity ($\eta_S = 0.89 \times 10^{-3}$ Pa s for water and $0.95 \times 10^{-3}$ Pa s for DMEM supplemented with calf serum and at T = 25 °C) [38]. The autocorrelation functions were interpreted using the cumulants and the CONTIN fitting procedure provided by the instrument software.

## III – Results and discussion

### III.1 – Effect of salt

Applications of NPs in biology require that they are stable in solutions containing high concentrations of proteins and salts, as well as in cell culture media. The particles put under scrutiny in this work were first tested with respect to changes in concentration, pH and ionic strength. With water as a solvent, stability was found up to c = 10 wt. %, and for pH = 5 – 11 [15]. Note that PPEG-coated particles were stable on a broader range, from pH = 1.5 to 10 [26]. Here, we illustrate the behavior as a function of the ionic strength $I_S$. The hydrodynamic diameters obtained by light scattering are displayed in Fig. 3 for $I_S = 0 – 1$ M. $I_S$ was adjusted using sodium (circles) and ammonium (squares) chloride. Fig. 3a shows that Cit–$\gamma$-Fe$_2$O$_3$ was destabilized by addition of both salts (at $I_S = 0.45$ M for NaCl and $I_S = 0.65$ M for NH$_4$Cl), whereas Cit–CeO$_2$ sustained ionic strength up to 1 M. The samples coated with pegylated oligomers (PPEG–CeO$_2$ and PPEG–$\gamma$-Fe$_2$O$_3$) and with ion-containing polymers (PAA$_{2K}$–CeO$_2$ and PAA$_{2K}$–$\gamma$-Fe$_2$O$_3$) were all stable up to $I_S = 1$ M. We have recently exploited this feature to control the aggregation of magnetic particles and to design superparamagnetic nanorods [36].

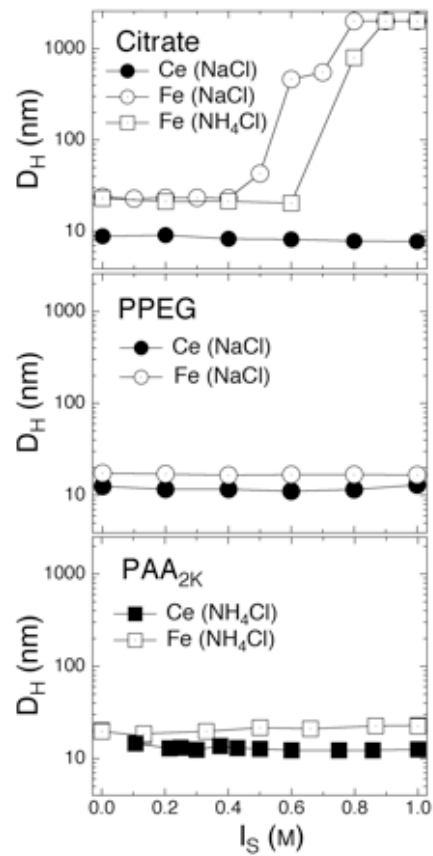

**Figure 3** : Hydrodynamic diameter for various inorganic nanoparticles dispersions (cerium oxide : full symbols; iron oxide : empty symbols) as a function of the ionic strength. $I_S$ was adjusted using sodium (circles) and ammonium (squares) chloride. Except for Cit–$\gamma$-Fe$_2$O$_3$, all other dispersions were stable at ionic strength up to 1 M.

### III.2 – Stability in cell culture medium

Figs. 4a and 4b show the time dependences of the Rayleigh ratio for the citrate-coated NPs Cit–CeO$_2$ and Cit–$\gamma$-Fe$_2$O$_3$ respectively. In these experiments, a few microliters of concentrated NP dispersions were mixed to 1 ml of freshly prepared culture medium. The time at which the mixing was made provided the origin on the abscissa (Figs. 4). As said in the experimental Section, the amount of added dispersion was calculated so as to cover molar concentrations [Ce] and [Fe] ranging from 0.1 mM to 10 mM. In the 6 experiments shown in Figs. 4, the Rayleigh ratio remained constant, but at values which did not scale with concentration (as expected from Eq. 2). For Cit–CeO$_2$ at [Ce] = 10 mM, a decrease of the intensity was even observed over the first 30 minutes of experiment. Further insights can be gained by comparing the levels of the scattering intensity with the predictions calculated from Eq. 2.

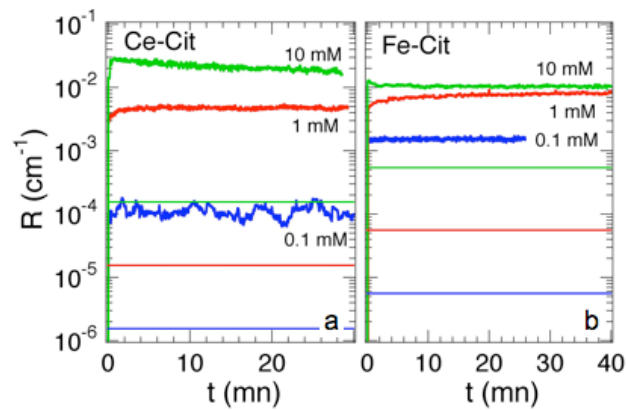

**Figure 4** : Time dependences of the Rayleigh ratio measured by static light scattering for Cit–CeO$_2$ (a) and Cit–γ-Fe$_2$O$_3$ (b) in DMEM supplemented with calf serum and antibiotics. The time at which nanoparticle dispersions were added to the culture medium provided the origin on the time axis. The concentrations targeted investigated were taken so as to cover the concentration ranges investigated in toxicology assays. Expressed in molar concentrations of metallic atoms, these concentrations ranged from {[Ce],[Fe]} = 0.1 – 30 mM. The color code for this viewgraph and in Figs. 4 – 8 is : blue for 0.1 mM, red for 1 mM, green for 10 mM and orange for 30 mM. The horizontal lines were calculated assuming that the particles remained in a dispersed state (from Eq. 2 and Table II). The color code was the same as for the data. The large scattering excess between the experimental and calculated values was interpreted as the signature of the destabilization of the citrate-coated particles.

These predictions were made using the results of Table II and assuming that : i) the particles remained in a dispersed state, and ii) the coupling constant K was identical in DI-water and in the culture medium. These estimations are the horizontal lines in Figs. 4 with the same color code as for the data (blue : 0.1 mM, red : 1 mM, green : 10 mM and orange : 30 mM). As a result, for Cit–CeO$_2$ and Cit–γ-Fe$_2$O$_3$, we have found that the actual scattering intensity was 10 to 1000 higher than that of dispersed particles. These findings are the signature that Cit–CeO$_2$ and Cit–γ-Fe$_2$O$_3$ NPs were destabilized at the contact of the culture medium. The destabilization was actually very rapid and intermediate values of the scattering could not be recorded. These outcomes also confirmed the direct visual observations of test tubes that showed high turbidity at the addition of particles. The experiments of Figs. 4 were restricted to short periods (< 1 hour), because at longer times the aggregates sedimented.

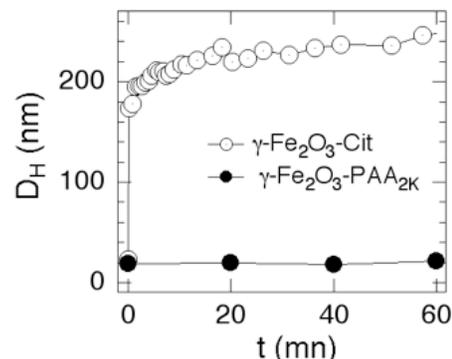

**Figure 5** : Time evolutions of the hydrodynamic diameter of Cit-γ-Fe$_2$O$_3$ and PAA$_{2K}$–γ-Fe$_2$O$_3$ dispersions ([Fe] = 1 mM) in DMEM supplemented with calf serum and antibiotics. The steep increase in D$_H$ at short times was interpreted as an indication of the destabilization of the dispersion. For the PAA$_{2K}$-coated particles, D$_H$ remained constant over the duration of the experiment, and over time longer than weeks.

In order to demonstrate further the destabilization of the nanosols, the hydrodynamic diameter of the Cit–γ-Fe$_2$O$_3$ dispersions ([Fe] = 1 mM) was monitored as a function of the time over a period of 1 hour (Fig. 5). At the mixing, D$_H$ exhibited a steep increase from its initial value D$_H$ = 23 nm to a value around 170 nm. The diameter evolved further and showed saturation around 250 nm. The increase in D$_H$ was also accompanied by a broadening of the size distribution. At longer times (> 1 hour), the large aggregates sedimented, resulting in the reduction of both scattering intensity and diameter. The data in Fig. 5 are in excellent agreement with those of the time evolution of the Rayleigh ratios for the same samples, both quantities indicating of destabilization of the particles.

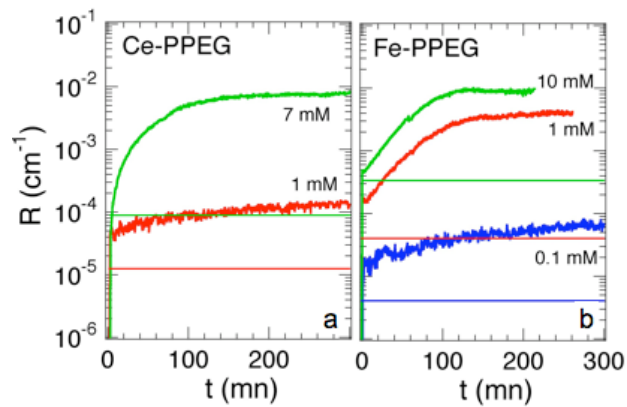

**Figure 6** : Same as in Figs. 4, but for PPEG–CeO$_2$ (a) and PPEG–γ-Fe$_2$O$_3$ (b) nanoparticles.

Figs. 6a and 6b display the Rayleigh ratios for the pegylated cerium and iron oxide NPs, PPEG–$CeO_2$ and PPEG–$\gamma$-$Fe_2O_3$ respectively. Data for 1 and 7 mM of cerium, and from 0.1 to 10 mM of iron oxide were collected over periods of 5 hours. For the 5 samples investigated, the same qualitative behavior was reported. Soon after the addition of the particles to the cell medium, the scattering intensities were of the order of those calculated for the disperse state. This state lasted a few minutes, and then the Rayleigh ratios started to increase, to finally saturate at much higher values ($\mathcal{R} = 10^{-2}$ cm$^{-1}$ for the most concentrated specimens). As for the citrate-coated particles, the excess scattering at steady state were 1 to 2 decades above the single particle predictions, indicating again a destabilization of the dispersions. As for the citrate-coated particles, this destabilization could be observed visually, since the suspending liquid became turbid after some time (> hours).

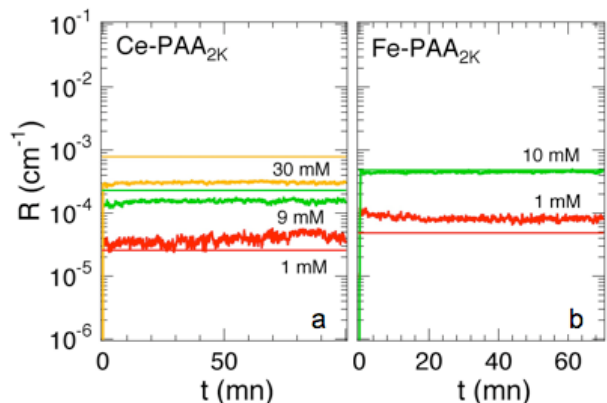

**Figure 7** : Same as in Figs. 4, but for $PAA_{2K}$–$CeO_2$ (a) and $PAA_{2K}$–$\gamma$-$Fe_2O_3$ (b) nanoparticles.

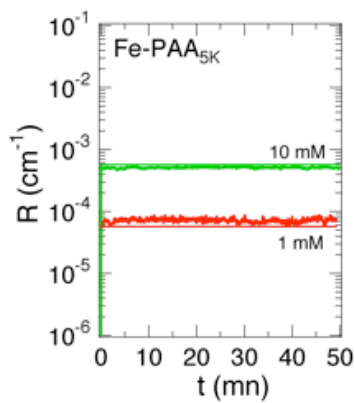

**Figure 8** : Same as in Figs. 4, but for $PAA_{5K}$–$\gamma$-$Fe_2O_3$ nanoparticles. The color code for this viewgraph is : red for 1 mM and green for 10 mM. The good agreement between the experimental and calculated results was interpreted as the indication of a superior stability in the cell culture medium.

As for the $PAA_{2K}$ and $PAA_{5K}$-coated NPs, the data in Figs. 7 and 8 show a constant level of scattering after the introduction of the particles into the culture medium. This property was observed at all concentrations investigated, up to [Ce] = 30 mM and [Fe] = 10 mM. Moreover, the scattering levels were in good agreement with the predictions made for disperse solutions. The agreement was indeed excellent for $PAA_{2K}$–$CeO_2$ at [Ce] = 1 mM, and for $PAA_{2K}$–$\gamma$-$Fe_2O_3$ and $PAA_{5K}$–$\gamma$-$Fe_2O_3$ at [Fe] = 1 and 10 mM. For the [Ce] = 30 mM specimen, we even observed a scattering intensity that was lower than the computed value. One possible reason for this was that at this concentration, interparticular (repulsive) interactions took place and lowered the intensity at the wave-vector of the experiment [30]. The remarkable stability of the $PAA_{2K}$-coated particles was further evidenced by the measurement of the hydrodynamic diameter as a function of time ([Fe] = 1 mM, lower curve in Fig. 5). There, $D_H$ remains unchanged over the duration of the experiment.

For the assessment of the colloidal stability over longer periods, the hydrodynamic diameters were determined one day and one week after the mixing. Tables III and IV list the $D_H$-values for the 19 cerium and iron oxide dispersions. Three stability trends are particularly noteworthy here : *i)* After a period of 1 week, the findings of Figs. 4 to 8 were corroborated : the citrate and PPEG-coated particles with $D_H$ comprised between 200 and 740 nm were sedimented at the bottom of the test tubes. In contrast, the poly(acrylic acid) coated particles were fully dispersed in the culture medium, with $D_H$ = 14 nm for $PAA_{2K}$–$CeO_2$ and 19 nm for $PAA_{2K}$–$\gamma$-$Fe_2O_3$. *ii)* The stability behavior of particles in cell culture medium depended more strongly on the coating than on the nature of the particle or the concentration. Particles with the same coating behave similarly. *iii)* When destabilization occured, as this is the case for the citrate and PEG-coated particles, the cluster sizes vary also with the initial concentration. As illustrated by the Cit–$CeO_2$ specimen in Table III, the higher the concentration, the larger the aggregates. This result is reminiscent from well-known mechanisms, such as the aggregation limited by the diffusion [39]. In summary, the most remarkable result obtained here was the excellent long term stability of the $PAA_{2K}$ and $PAA_{5K}$-coated particles.

## IV – Concluding remarks

In the present paper, we have shown that the coating plays an important role in the stability of aqueous dispersions of sub-10 nm cerium and iron oxide particles. Static and dynamic light scattering were performed as a function of time, for times comprised between 1 s to 1 week ($10^6$ s) after the introduction of particles into a conventional cell culture medium. A careful analysis of the data allows us to draw conclusions about the stabilization or destabilization mechanisms, as well as on the long term fate of the inorganic particles in these specific environments.

| coated–$CeO_2$ | [Ce] mM | $D_H$ (nm) DI-water | $D_H$ (nm) culture medium[1] | $D_H$ (nm) culture medium[2] |
|---|---|---|---|---|
| Cit–$CeO_2$ | 0.1<br>1<br>10 | 10 | 36<br>300<br>1150 | n.d.<br>330*<br>740* |
| PPEG–$CeO_2$ | 0.1<br>1<br>7 | 12 | 20<br>25<br>400 | 21<br>480<br>300* |
| $PAA_{2K}$–$CeO_2$ | 1<br>9<br>30 | 13 | 18<br>15<br>13 | n.d.<br>n.d.<br>14 |

**Table III** : Hydrodynamic diameters measured for coated nanoceria under 3 different solvent conditions : in de-ionizided water, in the cell culture medium after one day[1] and after one week[2]. The star (*) indicates dispersions that were diluted prior to the measurement so as to reduce absorption and multiple scattering.

| coated–$\gamma$-$Fe_2O_3$ | [Fe] mM | $D_H$ (nm) DI-water | $D_H$ (nm) culture medium[1] | $D_H$ (nm) culture medium[2] |
|---|---|---|---|---|
| Cit–$\gamma$-$Fe_2O_3$ | 0.1<br>1<br>10 | 23 | 184<br>220<br>375 | n.d.<br>200<br>n.d |
| PPEG–$\gamma$-$Fe_2O_3$ | 0.1<br>1<br>8 | 17 | 220<br>264<br>2380 | n.d.<br>n.d.<br>230* |
| $PAA_{2K}$–$\gamma$-$Fe_2O_3$ | 1<br>10 | 19 | 20<br>18 | 21<br>19 |
| $PAA_{5K}$–$\gamma$-$Fe_2O_3$ | 1<br>10 | 22 | 21<br>20 | 25<br>23 |

**Table IV** : Same data as in Table III, here for coated maghemite

According to DLVO theory, colloidal stability is insured when steric and electrostatic interactions are able to counterbalance the short-range van der Waals attractive interactions [40]. Following this approach, the destabilization observed with the citrate and PPEG-coated particles could arise from several mechanisms. A first mechanism would be related to the ionic strength gap that the particles underwent by switching from one solvent (DI-water, $I_S$ < 1 mM) to another (DMEM supplemented with calf serum, $I_S$ = 0.16 M). In this case, the range of the electrostatic interaction diminished through the decrease of the Debye screening length (from 10 nm to 0.7 nm), resulting in the onset of an aggregation process. Another possible mechanism for the destabilization could be the desorption of the adlayer through a competitive exchange with other components present in the culture medium. These components would then be the aminoacids in DMEM or the plasma proteins stemming from the calf serum. The multivalent counterions, such as calcium and magnesium present in DMEM could also play a crucial role. Changing the type and amount of species at the liquid-solid interface has the consequence to modify the range and ratio of the steric, electrostatic and van der Waals interactions between particles, and accelerate their uncontrolled clustering and final precipitation.

From the data shown in Fig. 3 ($D_H$ versus $I_S$), the citrate particles were destabilized by the addition of salts, but for ionic strengths above 0.45 M with NaCl and 0.65 M with $NH_4Cl$, i.e. higher than that of the growth medium. Therefore, the screening of the electrostatic interactions occurring at the mixing might not be the relevant mechanism here. We have also found that the particles behave very similarly in DMEM and in DMEM supplemented with the calf serum and antibiotics. We hence argue that the destabilization of the citrate and PPEG-coated particles probably started by the desorption of the coating through exchange with components present in the solvent. This mechanism is very rapid for the case of citrate (< 1s), slower for the PPEG-coated particles (hours).

From the above rationale, and regarding the excellent stability of the PAA-coated particles, it can been deduced that the poly(acrylic acid) chains were strongly attached to the particle surfaces. As already documented in Seghal et al. [24], we assume that this attachment occurred through the electrostatic adsorption of a few monomers (10 to 15 units for $PAA_{2K}$) [24] with the cationic charges on the surfaces, the remaining part of the chain being stretched towards the solvent. The anionic adlayers confer to the particles both the steric and electrostatic repulsions of charged brushes [41]. The combination of these two attributes (yielding the term electrosteric for the this type of repulsion) [22], together with the multi-point attachment of the chain are at the origin of the remarkable colloidal stability in high ionic strength solvents and in cell culture media. As a final comment, it is important to notice that hydrodynamic diameters of the PAA-coated particles were not altered in the cell culture medium, even after one week, indicating that there was no protein adsorption on the electrosteric brushes. These findings finally suggest that low molecular weight anionic polyelectrolytes represent a powerful alternative to conventional coating agents such as poly(ethylene oxide).

## Acknowledgements
We thank Olivier Sandre, Bruno Frka-Petesic, Régine Perzynski, Cyrille Richard, Nathalie Mignet, Jean-Paul Chapel, Hiba Sarrouj and Nathalie Luciani for numerous and fruitful discussions during the course of this work. The Laboratoire Physico-chimie des Electrolytes, Colloïdes et Sciences Analytiques and the Rhodia rare earths team at the Centre de Recherche d'Aubervilliers (Aubervilliers, France) are acknowledged for providing us with the nanoparticle dispersions. This research was supported in part by Rhodia (France), by the Agence Nationale de la Recherche, under the contract BLAN07-3_206866 and by the European Community through the project : "NANO3T—Biofunctionalized Metal and Magnetic Nanoparticles for Targeted Tumor Therapy", project number 214137 (FP7-NMP-2007-SMALL-1).